\documentclass[%
aps,
prl,
reprint,
superscriptaddress,
 amsmath,amssymb
]{revtex4-2}
\usepackage[T1]{fontenc}
\usepackage{graphicx}
\usepackage{dcolumn}
\usepackage{bm}
\usepackage[version=3]{mhchem}
\usepackage[dvipsnames]{xcolor}
\usepackage{xspace}
\usepackage[acronym,toc,shortcuts]{glossaries}
\usepackage{multirow}
\usepackage{verbatim}
\usepackage{placeins}

\begin{document}

\newcommand{\dSdOexp}{$\left(\frac{d\sigma}{d\Omega}\right)_{\mathrm{exp}}$}
\newcommand{\dSdOthry}{$\left(\frac{d\sigma}{d\Omega}\right)_{\mathrm{th}}$}

\newcommand{\piSD}{$\pi sd$\xspace}
\newcommand{\piSOneDFour}{$\pi s^{\,\,\,\,1}_{1/2} d^{\,\,\,\,4}_{3/2}$\xspace}
\newcommand{\piSTwoDThree}{$\pi s^{\,\,\,\,2}_{1/2} d^{\,\,\,\,3}_{3/2}$\xspace}

\hyphenation{VAMOS}
\hyphenation{MUGAST}
\hyphenation{AGATA}
\hyphenation{CATS}
\hyphenation{GANIL}

\preprint{APS/123-QED}
\title{Probing exotic cross-shell interactions at $N$\,=\,28 with single-neutron transfer on $^{47}$K}
\author{C.\,J.~Paxman}
    \affiliation{School of Maths and Physics, University of Surrey, Guildford, GU2 7XH, United Kingdom}
\author{A.~Matta}
    \affiliation{Université de Caen Normandie, ENSICAEN, CNRS/IN2P3, LPC Caen UMR6534, F-14000 Caen, France}
\author{W.\,N.~Catford}
    \affiliation{School of Maths and Physics, University of Surrey, Guildford, GU2 7XH, United Kingdom}
\author{G.~Lotay}
    \affiliation{School of Maths and Physics, University of Surrey, Guildford, GU2 7XH, United Kingdom}
\author{M.~Assi\'e}
    \affiliation{Universit\'{e} Paris-Saclay, CNRS/IN2P3, IJCLab, 91405 Orsay, France}
\author{E.~Clément}
    \affiliation{Grand Acc\'{e}l\'{e}rateur National d’Ions Lourds (GANIL), CEA/DRF-CNRS/IN2P3, Bvd Henri Becquerel, 14076 Caen, France}
\author{A.~Lemasson}
    \affiliation{Grand Acc\'{e}l\'{e}rateur National d’Ions Lourds (GANIL), CEA/DRF-CNRS/IN2P3, Bvd Henri Becquerel, 14076 Caen, France}
\author{D.~Ramos}
    \affiliation{Grand Acc\'{e}l\'{e}rateur National d’Ions Lourds (GANIL), CEA/DRF-CNRS/IN2P3, Bvd Henri Becquerel, 14076 Caen, France}
\author{N.\,A.~Orr}
    \affiliation{Université de Caen Normandie, ENSICAEN, CNRS/IN2P3, LPC Caen UMR6534, F-14000 Caen, France}
\author{F.~Galtarossa}
    \affiliation{Universit\'{e} Paris-Saclay, CNRS/IN2P3, IJCLab, 91405 Orsay, France}
\author{V.~Girard-Alcindor}
    \affiliation{Grand Acc\'{e}l\'{e}rateur National d’Ions Lourds (GANIL), CEA/DRF-CNRS/IN2P3, Bvd Henri Becquerel, 14076 Caen, France}
\author{J.~Dudouet}
    \affiliation{Universit\'{e} Claude Bernard Lyon 1, CNRS/IN2P3, IP2I Lyon, UMR 5822, F-69100 Villeurbanne, France}
\author{N.\,L.~Achouri}
    \affiliation{Université de Caen Normandie, ENSICAEN, CNRS/IN2P3, LPC Caen UMR6534, F-14000 Caen, France}
\author{D.~Ackermann}
    \affiliation{Grand Acc\'{e}l\'{e}rateur National d’Ions Lourds (GANIL), CEA/DRF-CNRS/IN2P3, Bvd Henri Becquerel, 14076 Caen, France}
\author{D.~Barrientos}
    \affiliation{CERN, CH-1211 Geneva 23 (Switzerland)}
\author{D.~Beaumel}
    \affiliation{Universit\'{e} Paris-Saclay, CNRS/IN2P3, IJCLab, 91405 Orsay, France}
\author{P.~Bednarczyk}
    \affiliation{The Henryk Niewodniczański Institute of Nuclear Physics, Polish Academy of Sciences, ul. Radzikowskiego 152, 31-342 Kraków, Poland}
\author{G.~Benzoni}
    \affiliation{INFN Sezione di Milano, I-20133 Milano, Italy}
\author{A.~Bracco}
    \affiliation{INFN Sezione di Milano, I-20133 Milano, Italy}
    \affiliation{Dipartimento di Fisica, Università di Milano, I-20133 Milano, Italy}
\author{L.~Canete}
    \affiliation{School of Maths and Physics, University of Surrey, Guildford, GU2 7XH, United Kingdom}
\author{B.~Cederwall}
    \affiliation{Department of Physics, KTH Royal Institute of Technology, SE-10691 Stockholm, Sweden}
\author{M.~Ciemala}
    \affiliation{The Henryk Niewodniczański Institute of Nuclear Physics, Polish Academy of Sciences, ul. Radzikowskiego 152, 31-342 Kraków, Poland}
\author{P.~Delahaye}
    \affiliation{Grand Acc\'{e}l\'{e}rateur National d’Ions Lourds (GANIL), CEA/DRF-CNRS/IN2P3, Bvd Henri Becquerel, 14076 Caen, France}
\author{D.\,T.~Doherty}
    \affiliation{School of Maths and Physics, University of Surrey, Guildford, GU2 7XH, United Kingdom}
\author{C.~Domingo-Pardo}
    \affiliation{Instituto de Física Corpuscular, CSIC-Universidad de Valencia, E-46071 Valencia, Spain}
\author{B.~Fern\'andez-Dom\'inguez}
    \affiliation{IGFAE and Dpt. de F\'{i}sica de Part\'{i}culas, Univ. of Santiago de Compostela, E-15758, Santiago de Compostela, Spain}
\author{D.~Fern\'andez} 
    \affiliation{IGFAE and Dpt. de F\'{i}sica de Part\'{i}culas, Univ. of Santiago de Compostela, E-15758, Santiago de Compostela, Spain}
\author{F.~Flavigny}
    \affiliation{Université de Caen Normandie, ENSICAEN, CNRS/IN2P3, LPC Caen UMR6534, F-14000 Caen, France}
\author{C.~Foug\`{e}res}
    \affiliation{Grand Acc\'{e}l\'{e}rateur National d’Ions Lourds (GANIL), CEA/DRF-CNRS/IN2P3, Bvd Henri Becquerel, 14076 Caen, France}
\author{G.~de~France}
    \affiliation{Grand Acc\'{e}l\'{e}rateur National d’Ions Lourds (GANIL), CEA/DRF-CNRS/IN2P3, Bvd Henri Becquerel, 14076 Caen, France}
\author{S.~Franchoo}
    \affiliation{Universit\'{e} Paris-Saclay, CNRS/IN2P3, IJCLab, 91405 Orsay, France}
\author{A.~Gadea}
    \affiliation{Instituto de Física Corpuscular, CSIC-Universidad de Valencia, E-46071 Valencia, Spain}
\author{J.~Gibelin}
    \affiliation{Université de Caen Normandie, ENSICAEN, CNRS/IN2P3, LPC Caen UMR6534, F-14000 Caen, France}
\author{V.~González}
    \affiliation{Departamento de Ingeniería Electrónica, Universitat de Valencia, Burjassot, Valencia, Spain}
\author{A.~Gottardo}
    \affiliation{Laboratori Nazionali di Legnaro, INFN, I-35020 Legnaro (PD), Italy}
\author{N.~Goyal}
    \affiliation{Grand Acc\'{e}l\'{e}rateur National d’Ions Lourds (GANIL), CEA/DRF-CNRS/IN2P3, Bvd Henri Becquerel, 14076 Caen, France}
\author{F.~Hammache}
    \affiliation{Universit\'{e} Paris-Saclay, CNRS/IN2P3, IJCLab, 91405 Orsay, France}
\author{L.\,J.~Harkness-Brennan}
    \affiliation{Oliver Lodge Laboratory, The University of Liverpool, Liverpool, L69 7ZE, UK}
\author{D.\,S.~Harrouz}
    \affiliation{Universit\'{e} Paris-Saclay, CNRS/IN2P3, IJCLab, 91405 Orsay, France}
\author{B.~Jacquot}
    \affiliation{Grand Acc\'{e}l\'{e}rateur National d’Ions Lourds (GANIL), CEA/DRF-CNRS/IN2P3, Bvd Henri Becquerel, 14076 Caen, France}
\author{D.\,S.~Judson}
    \affiliation{Oliver Lodge Laboratory, The University of Liverpool, Liverpool, L69 7ZE, UK}
\author{A.~Jungclaus}
    \affiliation{Instituto de Estructura de la Materia, CSIC, Madrid, E-28006 Madrid, Spain}
\author{A.~Kaşkaş}
    \affiliation{Department of Physics, Faculty of Science, Ankara University, 06100 Besevler - Ankara, Turkey}
\author{W.~Korten}
    \affiliation{Irfu, CEA, Université Paris-Saclay, F-91191 Gif-sur-Yvette, France}
\author{M.~Labiche}
    \affiliation{STFC Daresbury Laboratory, Daresbury, Warrington, WA4 4AD, UK}
\author{L.~Lalanne}
    \affiliation{Universit\'{e} Paris-Saclay, CNRS/IN2P3, IJCLab, 91405 Orsay, France}
    \affiliation{Grand Acc\'{e}l\'{e}rateur National d’Ions Lourds (GANIL), CEA/DRF-CNRS/IN2P3, Bvd Henri Becquerel, 14076 Caen, France}
\author{C.~Lenain}
    \affiliation{Université de Caen Normandie, ENSICAEN, CNRS/IN2P3, LPC Caen UMR6534, F-14000 Caen, France}
\author{S.~Leoni}
    \affiliation{INFN Sezione di Milano, I-20133 Milano, Italy}
    \affiliation{Dipartimento di Fisica, Università di Milano, I-20133 Milano, Italy}
\author{J.~Ljungvall}
    \affiliation{Universit\'{e} Paris-Saclay, CNRS/IN2P3, IJCLab, 91405 Orsay, France}
\author{J.~Lois\,-Fuentes}
    \affiliation{IGFAE and Dpt. de F\'{i}sica de Part\'{i}culas, Univ. of Santiago de Compostela, E-15758, Santiago de Compostela, Spain}
\author{T.~Lokotko}
    \affiliation{Université de Caen Normandie, ENSICAEN, CNRS/IN2P3, LPC Caen UMR6534, F-14000 Caen, France}
\author{A.~Lopez-Martens}
    \affiliation{Universit\'{e} Paris-Saclay, CNRS/IN2P3, IJCLab, 91405 Orsay, France}  
\author{A.~Maj}
    \affiliation{The Henryk Niewodniczański Institute of Nuclear Physics, Polish Academy of Sciences, ul. Radzikowskiego 152, 31-342 Kraków, Poland}
\author{F.\,M.~Marqu\'{e}s}
    \affiliation{Université de Caen Normandie, ENSICAEN, CNRS/IN2P3, LPC Caen UMR6534, F-14000 Caen, France}
\author{I.~Martel}
    \affiliation{Departamento de Ciencias Integradas, Universidad de Huelva, Calle Dr.~Cantero Cuadrado, 6, 21004 Huelva, Spain}
\author{R.~Menegazzo}
    \affiliation{INFN Sezione di Padova, I-35131 Padova, Italy}
\author{D.~Mengoni}
    \affiliation{INFN Sezione di Padova, I-35131 Padova, Italy}
    \affiliation{Dipartimento di Fisica e Astronomia dell'Università di Padova, I-35131 Padova, Italy}
\author{B.~Million}
    \affiliation{INFN Sezione di Milano, I-20133 Milano, Italy}
\author{J.~Nyberg}
    \affiliation{Department of Physics and Astronomy, Uppsala University, SE-75120 Uppsala, Sweden}
\author{R.\,M.~Pérez-Vidal}
    \affiliation{Instituto de Física Corpuscular, CSIC-Universidad de Valencia, E-46071 Valencia, Spain}
    \affiliation{Laboratori Nazionali di Legnaro, INFN, I-35020 Legnaro (PD), Italy}
\author{L.~Plagnol}
    \affiliation{Université de Caen Normandie, ENSICAEN, CNRS/IN2P3, LPC Caen UMR6534, F-14000 Caen, France}
\author{Zs.~Podolyák}
    \affiliation{School of Maths and Physics, University of Surrey, Guildford, GU2 7XH, United Kingdom}
\author{A.~Pullia}
    \affiliation{INFN Sezione di Milano, I-20133 Milano, Italy}
    \affiliation{Dipartimento di Fisica, Università di Milano, I-20133 Milano, Italy}
\author{B.~Quintana}
    \affiliation{Laboratorio de Radiaciones Ionizantes, Departamento de Física Fundamental, Universidad de Salamanca, E-37008 Salamanca, Spain}
\author{D.~Regueira-Castro}
    \affiliation{IGFAE and Dpt. de F\'{i}sica de Part\'{i}culas, Univ. of Santiago de Compostela, E-15758, Santiago de Compostela, Spain}
\author{P.~Reiter}
    \affiliation{Institut für Kernphysik, Universität zu Köln, Zülpicher Str. 77, D-50937 Köln, Germany}
\author{M.~Rejmund}
    \affiliation{Grand Acc\'{e}l\'{e}rateur National d’Ions Lourds (GANIL), CEA/DRF-CNRS/IN2P3, Bvd Henri Becquerel, 14076 Caen, France}
\author{K.~Rezynkina}
    \affiliation{Université de Strasbourg, CNRS, IPHC UMR 7178, F-67000 Strasbourg, France}
    \affiliation{INFN Sezione di Padova, I-35131 Padova, Italy}
\author{E.~Sanchis}
    \affiliation{Departamento de Ingeniería Electrónica, Universitat de Valencia, Burjassot, Valencia, Spain}
\author{M.~Şenyiğit}
    \affiliation{Department of Physics, Faculty of Science, Ankara University, 06100 Besevler - Ankara, Turkey}
\author{N.~de~S\'er\'eville}
    \affiliation{Universit\'{e} Paris-Saclay, CNRS/IN2P3, IJCLab, 91405 Orsay, France}
\author{M.~Siciliano}
    \affiliation{Laboratori Nazionali di Legnaro, INFN, I-35020 Legnaro (PD), Italy}
    \affiliation{Irfu, CEA, Université Paris-Saclay, F-91191 Gif-sur-Yvette, France}
    \affiliation{Physics Division, Argonne National Laboratory, Lemont (IL), United States}
\author{D.~Sohler} 
    \affiliation{Institute for Nuclear Research, Atomki, 4001 Debrecen, P.O. Box 51, Hungary}
\author{O.~Stezowski}
    \affiliation{Universit\'{e} Claude Bernard Lyon 1, CNRS/IN2P3, IP2I Lyon, UMR 5822, F-69100 Villeurbanne, France}
\author{J.-C.~Thomas}
    \affiliation{Grand Acc\'{e}l\'{e}rateur National d’Ions Lourds (GANIL), CEA/DRF-CNRS/IN2P3, Bvd Henri Becquerel, 14076 Caen, France}
\author{A.~Utepov}
    \affiliation{Grand Acc\'{e}l\'{e}rateur National d’Ions Lourds (GANIL), CEA/DRF-CNRS/IN2P3, Bvd Henri Becquerel, 14076 Caen, France}
\author{J.\,J.~Valiente-Dobón}
    \affiliation{Laboratori Nazionali di Legnaro, INFN, I-35020 Legnaro (PD), Italy}
\author{D.~Verney}
    \affiliation{Universit\'{e} Paris-Saclay, CNRS/IN2P3, IJCLab, 91405 Orsay, France}
\author{M.~Zielińska}
    \affiliation{Irfu, CEA, Université Paris-Saclay, F-91191 Gif-sur-Yvette, France}

\date{\today}
             
\begin{abstract}

We present the first measurement of the $^{47}$K($d,p\gamma$)$^{48}$K transfer reaction, performed in inverse kinematics using a reaccelerated beam of $^{47}$K. The level scheme of $^{48}$K has been greatly extended, with nine new bound excited states identified and spectroscopic factors deduced. Uniquely, the $^{47}$K($d,p$) reaction gives access to nuclear states that are sensitive to the interaction of protons and neutrons in the widely-spaced $1s$ and $fp$ orbitals, respectively. Detailed comparisons with SDPF-U and SDPF-MU shell-model calculations reveal a number of discrepancies between theory and experiment. Intriguingly, a systematic overestimation of spectroscopic factors and a poor reproduction of the energies for 1$^-$ states suggests that the mixing between the \piSOneDFour{} and \piSTwoDThree{} proton configurations in $^{48}$K is not correctly described using current interactions, challenging our description of light nuclei around the $N$\,=\,28 island of inversion.

\end{abstract}

\maketitle

The nuclear shell model has long been established as the central theoretical approach to studying the structure of nuclei across the chart of isotopes~\cite{Mayer1949_ShellsInNuclei, HaxelJensenSuess1949_SpinOrbSplit, BrownWildenthal1988_ReviewShellModel, Brown2022_ShellModelTowardsDripLines}. At its core is the premise that complex many-body nuclear systems may be described as protons and neutrons occupying orbitals of discrete energies, organized in shells. Specific nuclear states are then obtained by computational diagonalizations of the nuclear Hamiltonian in a many-body model space consisting of one or more shells. This approach has been extremely effective in reproducing experimental observations, including excitation energies of nuclear states, spin-parity assignments, and the emergence~\cite{Steppenbeck2013_N34MagicNUmberInCalcium, Lalanne2023_N16magicity} or weakening~\cite{Sorlin2013_n28Evolution} of ``magic numbers'' (particularly stable configurations of nucleons). However, these models are partly phenomenological, requiring experimental data to refine the interactions between orbitals, and, as such, are less well suited to describing exotic, `cross-shell' interactions, for which experimental data is sparse – these occur between orbitals that have a significant separation in energy. Consequently, there is now a critical need for experimental studies of the $\pi sd$$-\nu fp$ cross-shell interactions, which will strongly influence the structure of neutron-rich nuclei from $^{42}$Si to $^{60}$Ca.

Previous studies of the $\pi sd$$-\nu fp$ interaction have been very successful, establishing the emergence of new magic numbers~\cite{Steppenbeck2013_N34MagicNUmberInCalcium} and observing a critical weakening of $p$-wave spin-orbit splitting~\cite{Gaudefroy2006_46Ardp_Original, Gaudefroy2006_46Ardp_SignoracciComment, Gaudefroy2006_46Ardp_Reply}. Of these, the interpretation of $^{46}$Ar($d,p$) is not straightforward due to the ground state structure of $^{46}$Ar being poorly understood~\cite{Mengoni2010_46Ar-lifetimeMeas, Calinescu2016_46ArCoulEx,Nowak2016_46Ar-twoNtransfer}. For the neighbouring $^{47}$K nucleus, however, the ground state structure is well-understood due to a previous measurement of the magnetic moment~\cite{Kreim2014_LaserSpec_48KGroundState}. Specifically, $^{47}$K$_{\textrm{g.s.}}$ has an unusual proton configuration, \piSOneDFour, whereas its neighbour nuclei, $^{46}$K$_{\textrm{g.s.}}$ and $^{48}$K$_{\textrm{g.s.}}$, both have primarily \piSTwoDThree{} configurations~\cite{Papuga2014}. This makes $^{47}$K$_{\textrm{g.s.}}$ a solid foundation for transfer reaction studies, whilst providing the first access to the $\pi s_{1/2}$$-\nu fp$ interaction by this method. The level structure of $^{48}$K, however, is largely unknown~\cite{Krolas_2011}, and current shell-model calculations fail to reproduce the experimentally determined spin and parity (1$^{-}$) of the ground state. It seems that the mixing between the proton configurations \piSOneDFour{} and \piSTwoDThree{}, which coexist at near-degenerate energies, may play a key role in the structure of $^{48}$K.

\begin{figure*}
	\centering
    \includegraphics[width=\linewidth]{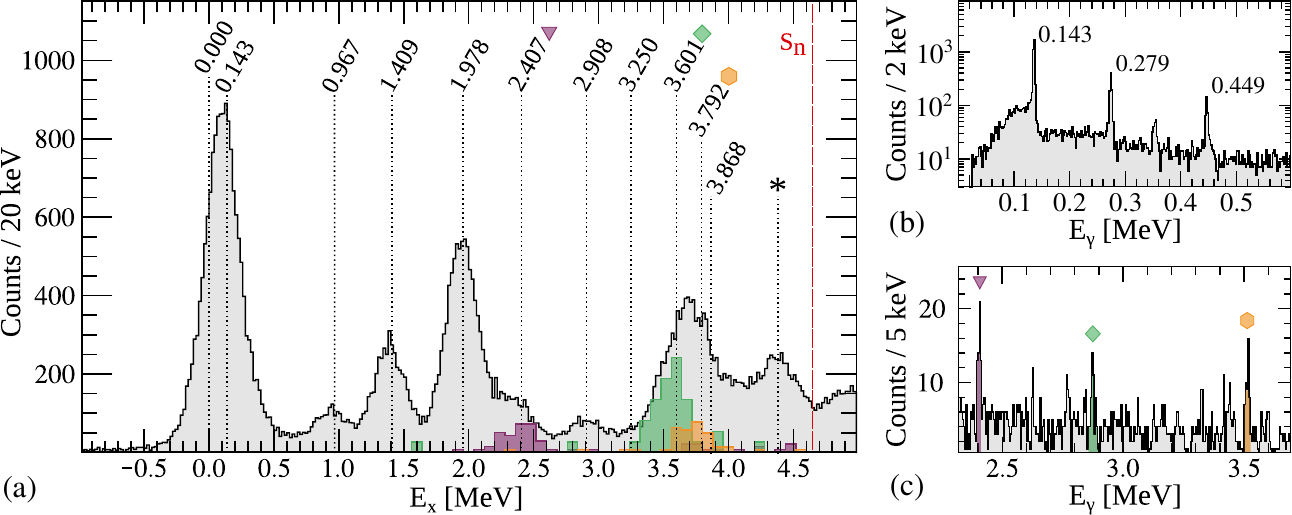}
	\caption{(Color online) (a) Full reconstructed excitation energy spectrum of $^{48}$K in black, with the neutron separation energy, $S_n = 4.644$~MeV, indicated. Peaks are labeled with their energies as determined by $\gamma$-ray spectroscopy. Three excitation energy spectra in coincidence with three observed $\gamma$-rays (scaled by branching ratio and efficiency) are shown in colour, showcasing how otherwise-unresolved peaks can be resolved. The peak marked with an asterisk represents an unresolved multiplet -- see text. (b) Gamma-decays observed with AGATA over the energy range 0.0$-$0.6 MeV (cf.~Table~\ref{tableOfStates}). The small peak at 0.357(1)~MeV corresponds to the decay of the first excited state in $^{47}$K (slightly shifted due to its 1~ns lifetime~\cite{Burrows2007_NDS_47K}) which is populated here by the decay of neutron-unbound states in $^{48}$K. (c) Gamma-decays observed with AGATA over the energy range 2.3$-$3.7 MeV. Symbols indicate the $\gamma$-rays (2.407, 2.872 and 3.516~MeV) used to construct the coloured excitation spectra.}
	\label{fExEg}
\end{figure*}

In this Letter, we present a detailed spectroscopic study of the nucleus $^{48}$K, in which excited states up to and above the neutron-emission threshold energy (\ce{S_n} = 4.644~MeV~\cite{NuclearDataSheets_Mass48}) were populated via the $^{47}$K($d,p\gamma$) transfer reaction in inverse kinematics (i.e.~a heavy beam incident on a light target). This selective reaction mechanism allows for the preferential population of single-particle levels in $^{48}$K, thereby isolating states built upon the \piSOneDFour ground state of $^{47}$K, with a neutron in the $\nu p_{3/2}$, $\nu p_{1/2}$ or hitherto-unexamined $\nu f_{5/2}$ orbitals. The single-particle structure is examined by extracting spectroscopic factors -- that is, the overlap of initial $^{47}$K$_{g.s} + d$ and final $^{48}$K$ + p$ state wavefunctions. Moreover, in probing such states, this work represents a stringent test of modern shell-model calculations, requiring a need to account for both the neutron wavefunction and the mixing between the two proton configurations. In this regard, the unique combination~\cite{MugastAgataVamos_NIMA} of the AGATA high-efficiency $\gamma$-ray tracking array~\cite{Agata_NIMA,Clement2017_Agata1piAtGanil}, the advanced MUGAST silicon array~\cite{Assie2020_MUGASTatGANIL}, and the VAMOS++ magnetic spectrometer~\cite{Rejmund2011_VAMOS++}, was of critical importance. In particular, it provided the background-free, ultra-high-resolution performance required to extract nuclear properties from excited states in the odd-odd nucleus $^{48}$K, which are closely spaced within the energy range $E_x$ = 0 – 4~MeV.

A thick graphite target was bombarded with a 60 MeV/nucleon primary beam of $^{48}$Ca, in order to produce a 7.7 MeV/nucleon beam of unstable $^{47}$K ions, delivered by the SPIRAL1+ facility at GANIL~\cite{Chauveau2023_SPIRAL1}. This >99.9\% pure secondary beam of $^{47}$K, of intensity $\sim 5 \times 10^{5}$~pps, was directed onto a 0.31(2)~mg/cm$^{2}$ thick self-supporting \ce{CD_2} target. Prompt $\gamma$ rays were detected using the state-of-the-art AGATA tracking array~\cite{Stezowski2023_AgataSoftware,Boston2023_AgataPSA,Crespi2023_AgataTracking,LopezMartens2004_TrackAlgComparison}, which in this instance consisted of 12 triple-cluster HPGe detectors mounted 18~cm from the target, covering angles $\theta_{\textrm{lab}}$$\sim$130\,–\,160$^\circ$~\cite{MugastAgataVamos_NIMA}. Ejected protons were detected at backward angles ($\theta_{\textrm{lab}}$$\sim$104\,–\,156$^\circ$) with the MUGAST double-sided silicon strip detector array, while beam-like heavy-ion recoils were transmitted to the focal plane of the VAMOS++ magnetic spectrometer (providing effective rejection of events resulting from reactions on carbon). Beam normalisation was achieved by monitoring incoming $^{47}$K particles with a single CATS multiwire proportional counter~\cite{OttiniHustache1999_CATS} and by studying elastically scattered deuterons with a silicon detector, placed downstream of the target position. For proton energy calibrations, a triple-$\alpha$ source consisting of $^{239}$Pu, $^{241}$Am and $^{244}$Cm was employed. The $^{48}$K excitation energy ($E_x$) was reconstructed from the energy and the angle of the protons, requiring a coincidence with an ion at the VAMOS++ focal plane. Whilst the resolution for the kinematically reconstructed excitation energy was $\textrm{FWHM}\sim$330~keV, the coincident $\gamma$-ray energy measurement allowed for a much higher effective resolution to be achieved. Gamma-ray energy and efficiency calibrations were carried out using a standard $^{152}$Eu source, extrapolated to high energies through simulation~\cite{Farnea2010_AgataSimulations}. Doppler correction of $\gamma$-ray energies was performed using velocities calculated event-by-event, from the observed proton energies and angles. The Doppler-corrected $\gamma$-rays had a resolution of $\textrm{FWHM}\sim$7~keV at 1.8~MeV, allowing for states that are unresolved in the particle-only spectrum to be separated in the particle-$\gamma$ coincident spectrum, detailed below in Fig.~\ref{fExEg}.

An elastic scattering normalisation value, $N_{d}$ (the product of the number of deuterons in the target and the integrated flux of the incoming beam), was determined by comparing elastically scattered deuteron data with optical model calculations. In this case, the optical model calculation was performed using the code FRESCO~\cite{fresco} and the potential of Daeknick, Childs and Vrcelj~\cite{Optical_DCV}, although we note that DWUCK4~\cite{dwuck} produces entirely consistent results. Observed ($d,p$) proton events were reconstructed using the \textit{nptool} framework~\cite{Matta_2016_nptool}, which also allowed a consistent efficiency determination for protons using realistic \textsc{Geant4} Monte Carlo simulations~\cite{Agostinelli2003_GEANT4}. The factor $N_{d}$ was then applied to the observed angular distribution of these protons in order to obtain differential cross sections ($d\sigma/d\Omega$). For the extraction of spectroscopic factors, $S$, differential cross sections were compared with theoretical calculations determined from the code TWOFNR~\cite{TWOFNR}, using the Koning-Delaroche global optical potential~\cite{KoningDelarocheRef} and Johnson-Tandy adiabatic model~\cite{JohnsonTandyRef,JohnsonSoperRef}. In this analysis, the statistical error in extracting spectroscopic factors from differential cross sections is $<$10$\%$~\cite{Paxman2024_Thesis}. However, the overall systematic uncertainty arising from reaction model limitations is expected to be $\sim$20\%. This is illustrated in the analysis of Ref.~\cite{JennyLee20percent}, which also demonstrates that this technique, as discussed in Ref.~\cite{Matta2019_29Mg}, provides spectroscopic factors that can be compared directly to the shell model values. Shell model calculations were performed using the NuShellX code~\cite{Brown2014_NuShellX} to compute the first 40 states of each spin. Two state-of-the-art interactions were used, SDPF-U~\cite{Nowacki2009_SDPF-U} and SDPF-MU~\cite{Otsuka2010_SDPF-MU_Vmu,Utsuno2012_SDPF-MU}, as they are well-suited to describing nuclei in the region $N$>28, $Z$<20. Crucially, while the cross-shell $\pi sd$$-\nu fp $ component of the SDPF-U interaction is phenomenological in nature, in SDPF-MU this component is instead derived from first principles.

\begin{table}
\caption{Properties of states observed in $^{48}$K. Excited state energies were determined using $\gamma$-ray transitions observed from states populated directly in ($d,p$) or in cascade decays. Associated branching ratios (BR) are listed. Experimental spectroscopic factors $S_\textrm{exp}$ are shown in comparison with shell-model calculations using the SDPF-MU~\cite{Otsuka2010_SDPF-MU_Vmu,Utsuno2012_SDPF-MU} ($S_{\textrm{MU}}$) and SDPF-U~\cite{Nowacki2009_SDPF-U}  ($S_{\textrm{U}}$) interactions. Upper limits on $S_\textrm{exp}$ have been established from the non-observation of proton peaks. See supplemental material for a full level scheme diagram.}
\begin{tabular}{cc|cc|cccc}
\hline	\hline
\ce{E_{x}} [MeV]  &	J$^\pi$	&	\ce{E_{\gamma}} [MeV] &	 BR	& 	$n\ell_{j}$	&	$S_{\textrm{exp}}$\footnotemark[1]	&	$S_{\textrm{MU}}$ &	$S_{\textrm{U}}$	\\ \hline	\hline	
0.000\phantom{(1)}	&	$1^{-}$ &  ---	  &	---	      	 & 2$p_{3/2}$ &	0.24(5)	&	0.40	&	0.21	\\	\hline	
0.143(1)	&	$2^{-}$ &  0.143(1) &	$\sim$100 	& 2$p_{3/2}$ &	0.42(8)	&	0.86	&	0.84	\\	\hline	
\phantom{\footnotemark[2]}0.279(1)\footnotemark[2]   &  $2^{-}$  &	0.279(1) &	$\sim$100 	& 2$p_{3/2}$ &	$<$0.03 	&	0.01 	&	0.05     \\	\hline	
\phantom{\footnotemark[2]}0.728(3)\footnotemark[2]   &  $3^{-}$  &	0.449(2) &	$\sim$100 	&   1$f_{7/2}$ &	$<$0.04 	&	0.06 	&	0.05     \\	\hline	
0.967(2)	&	$0^{-}$ &  0.967(2) &	$\sim$100  & 2$p_{1/2}$   &	0.26(5)	&	0.40	&	0.38	\\	\hline	
1.409(3)	&	$1^{-}$ &  1.130(3) &   10(2)	   & 2$p_{3/2}$ &	0.24(5)	&	0.35	&	0.54	\\		
	        &           &  1.266(2) &   63(2)      & 	            &		    &		    &		    \\		
	        &           &  1.409(3) &   28(1)      & 		        &		    &		    &		    \\  \hline	
1.978(4)	&	$1^{-}$	&  1.010(4) &	\,\,\,5(1) & 2$p_{1/2}$   &	0.50(10)&	0.88	&	0.84	\\		
	        &           &  1.836(3) &   93(2)      & 	            &		    &		    &		    \\		
	        &           &  1.979(3) &   \,\,\,2(1) & 	            &		    &		    &		    \\  \hline	
2.407(6)	&   $0^{-}$	&  0.997(4) &	33(2)	   & 2$p_{1/2}$   &	0.34(7)	&	0.56	&	0.58	\\		
	        &           &  2.407(5) &   67(7)      &                &		    &		    &		    \\  \hline	
2.908(8)	&	$2^{-}$ &  2.765(7) &	$\sim$100  & 2$p_{3/2}$ &  0.023(5)	&		    &		    \\		
	        &           &           &		       & 1$f_{5/2}$  &  0.06(1)  &	--- 	&	--- 	\\  \hline	
3.250(6)	& $(3^{-})$ &  2.971(4) &	$\sim$100  & 1$f_{5/2}$  &	0.06(1)	&	0.11	&	    	\\	\hline	
3.601(8)	&	$2^{-}$	&  2.193(3) &	15(4)	   & 1$f_{5/2}$  &	0.34(7)	&	0.47	&	0.50	\\		
	        &           &  2.872(7) &   38(7)      & 	            &		    &			&		    \\		
	        &           &  3.325(4) &   22(5)      & 	            &		    &			&		    \\		
	          &           &  3.458(7) &   12(4)      & 	              &		      &			  &		      \\		
	          &           &  3.598(7) &   12(4)      & 	              &		      &			  &		      \\  \hline	
3.792(8)	& $(3^{-})$ &  3.063(2) &	12(4)	   & 1$f_{5/2}$  &	0.16(3)	&	0.33	&	0.39	\\		
	        &           &  3.516(7) &   88(4)      & 	            &  		    &		    &		    \\  \hline	
3.868(7)	& $(2^{-})$	&  3.727(6) &	59(9)	   & 1$f_{5/2}$  &	0.14(3)	&	0.18    &	0.21    \\		
	        &           &  3.865(8) &   41(9)      & 		        &		    &			&		    \\  \hline	
\hline
\vspace{-0.35cm}\footnotetext[1]{Systematic uncertainty of 20\% has been applied, see text.}
\footnotetext[2]{State inferred from the $\gamma$ decays of higher-energy excited states.}
\end{tabular}
\label{tableOfStates}
\end{table}

Table~I presents the excitation energies, spin-parity assignments and spectroscopic factors of states in $^{48}$K observed in this work, together with a comparison to current shell-model calculations, while Fig.~\ref{fExEg} displays the excitation energy spectrum and coincident $\gamma$-ray spectrum observed in this work. With the exception of states up to 0.8~MeV~\cite{Krolas_2011}, all other states in $^{48}$K are reported here for the first time. It is important to note that all levels observed in this study must have some wavefunction component of the unusual \piSOneDFour{} structure which characterises $^{47}$K$_{\textrm{g.s.}}$ -- we do not observe the three states above 0.8~MeV found by Kr\'olas {\it et al.}~\cite{Krolas_2011}, as they have a structure based on the \piSTwoDThree{} configuration.

\begin{figure}
	\centering
    \includegraphics[width=\linewidth]{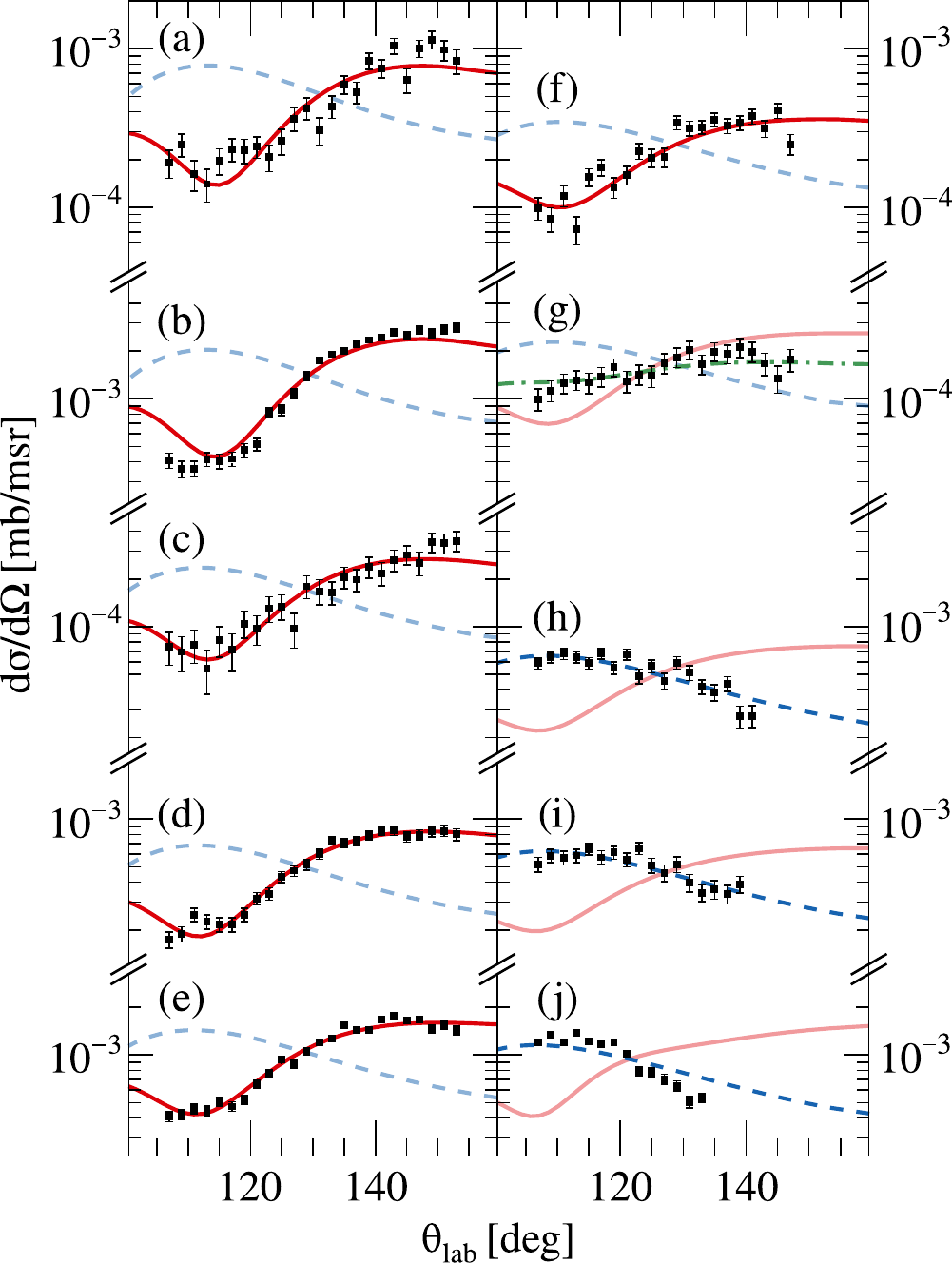}
	\caption{Experimental differential cross sections for states observed in this work, with the scaled theoretical differential cross sections for $\ell_p$=1 ($\ell_p$=3) transfer in solid red (dashed blue) lines: (a)~Ground state, (b)~0.143~MeV, (c)~0.967~MeV, (d)~1.409~MeV, (e)~1.978~MeV, (f)~2.407~MeV, (g)~2.908~MeV, (h)~3.601~MeV, (i)~3.792 \& 3.868~MeV, (j)~High energy multiplet. Only in the case of (g) was both $\ell$=1 and $\ell$=3 clearly required, with the best-fit shown in dot-dash green.}
	\label{fCSfits}
\end{figure}

In considering Fig.~\ref{fExEg}, several states in $^{48}$K, such as the ground state and the first excited state (0.143 MeV), are unresolved. However, using the excitation energies obtained via $\gamma$-ray spectroscopy, a function was constructed which fit every state simultaneously, reproducing the excitation energy spectra well, across all angles. Hence, individual angular distributions were extracted for each state. Fig.~\ref{fCSfits} presents the $d\sigma/d\Omega$ of observed states, alongside theoretical calculations for $\ell=1,3$ transfer.
As shown in Figs.~\ref{fCSfits}(a) and (b), both the ground and first excited states in $^{48}$K are well described by $\ell=1$ transfer, in agreement with the previous 1$^{-}_1$ and 2$^{-}_1$ assignments~\cite{Kreim2014_LaserSpec_48KGroundState, Krolas_2011}. Beyond the two lowest energy states, four strong peaks are observed in Fig.~\ref{fExEg} at 0.967, 1.409, 1.978 and 2.407~MeV. An angular distribution analysis of these peaks, shown in Figs.~\ref{fCSfits}(c)-(f), reveals pure $\ell$ = 1 characters, indicating spin-parity assignments of 0$^{-}$, 1$^{-}$ or 2$^{-}$ only. Furthermore, the $\gamma$-decay pathways of these four states (see Table~\ref{tableOfStates}) suggest spin-parities of 0$^{-}$, 1$^{-}$, 1$^{-}$ and 0$^{-}$, respectively. This is in agreement with the predicted ordering of these states in both shell-model calculations. As such, we adopt these assignments here.

In contrast, at higher energies ($E_x$ $>$ 3~MeV) all states appear to be populated via pure $f-$wave ($\ell=$ 3) transfer -- see Figs.~\ref{fCSfits}(h)-(i) -- indicating $J^{\pi}$ assignments of 2$^{-}$ or 3$^{-}$, assuming transfer to the empty $f_{5/2}$ orbital. Of these, the 3.601~MeV state is the most strongly populated, with $\gamma$ decays to 1$^{-}$, 2$^{-}$ and 3$^{-}$ levels suggesting it is 2$^{-}$ in character. Additionally, both shell-model calculations predict that the state populated with the largest pure $f-$wave spectroscopic factor is a 2$^{-}$ excited state. Therefore, we assign the 3.601~MeV state in $^{48}$K as 2$^{-}$, associated with the strongest 2$^{-}$ state in each calculation. Of the two close-lying states at 3.792~MeV and 3.868~MeV, the former is found to exhibit $\gamma$-ray transitions to the $2^-$, 0.279~MeV and the $3^-$, 0.728~MeV states, while the latter decays to the $2^-$, 0.143~MeV and the $1^-$ ground states. In this light, we favour 3$^{-}$ and 2$^{-}$ assignments for 3.792~MeV and 3.868~MeV, respectively. We associate these levels with the 3$^-_5$, 3.382~MeV and 2$^-_7$, 3.522~MeV states in SDPF-MU and the 3$^-_5$, 3.810~MeV and 2$^-_5$, 3.694~MeV states in SDPF-U, as these represent the closest matches in terms of energy and predicted spectroscopic factor. In addition, a relatively weakly populated $\ell=$ 3 state at 3.250~MeV is found to decay via a 2.971~MeV $\gamma$ ray to the second excited 2$^-$ state. As we do not observe the energetically favoured transition to the 1$^-$ ground state, and based on comparisons with the shell model, we tentatively assign this state as 3$^{-}$. We expect a number of levels to contribute to the peak labelled by an asterisk in Fig.~\ref{fExEg}. As such, an angular distribution analysis was performed by considering this high-energy multiplet to be a single broad peak. Whilst pure $\ell=3$ transfer reproduces the data well -- see Fig.~\ref{fCSfits}(j) -- definitive spin-parity assignments could not be determined. Consequently, we simply report here an upper (lower) limit on the overall spectroscopic factor for this region $\sim$4~MeV of 0.45 $\pm$ 0.09 (0.32 $\pm$ 0.07). Finally, for states above the neutron threshold, we attribute an integrated $\ell=3$ spectroscopic factor of $\sim$0.2. 

Intriguingly, as can be seen in Fig.~\ref{fCSfits}(g), we observe an excited state at 2.908~MeV that is not well reproduced by either a pure $\ell=1$ or $\ell=3$ distribution, and requires a mixed configuration, indicating a 2$^{-}$ assignment. This is of particular significance as no such state is predicted by either shell-model calculation. Here, we extract $p_{3/2}$ and $f_{5/2}$ spectroscopic factors of 0.023(5) and 0.06(1) for the 2.908~MeV excited level. Furthermore, given its definitive $f-$wave component, we note that the presently observed 2.908~MeV state provides an excellent marker for the location of the $f_{5/2}$ shell in this region of the nuclear chart. Using this, together with the results for spectroscopic factors and spins presented above, we extract neutron single-particle energies for $f_{5/2}$, $p_{1/2}$, $p_{3/2}$ and $f_{7/2}$ of $-0.6$, $-2.71$, $-4.19$ and $-8.37$~MeV, respectively. We note that these results in $^{48}$K lie in between those of $^{49}$Ca ($-1.15$, $-3.13$, $-5.15$ and $-9.95$~MeV) and $^{47}$Ar ($-0.1$, $-2.42$, $-3.55$ and $-8.02$~MeV) -- cf.\,Fig~3 in Ref~\cite{Gaudefroy2006_46Ardp_Original}.

\begin{figure}[b]
	\centering
    \includegraphics[width=\linewidth]{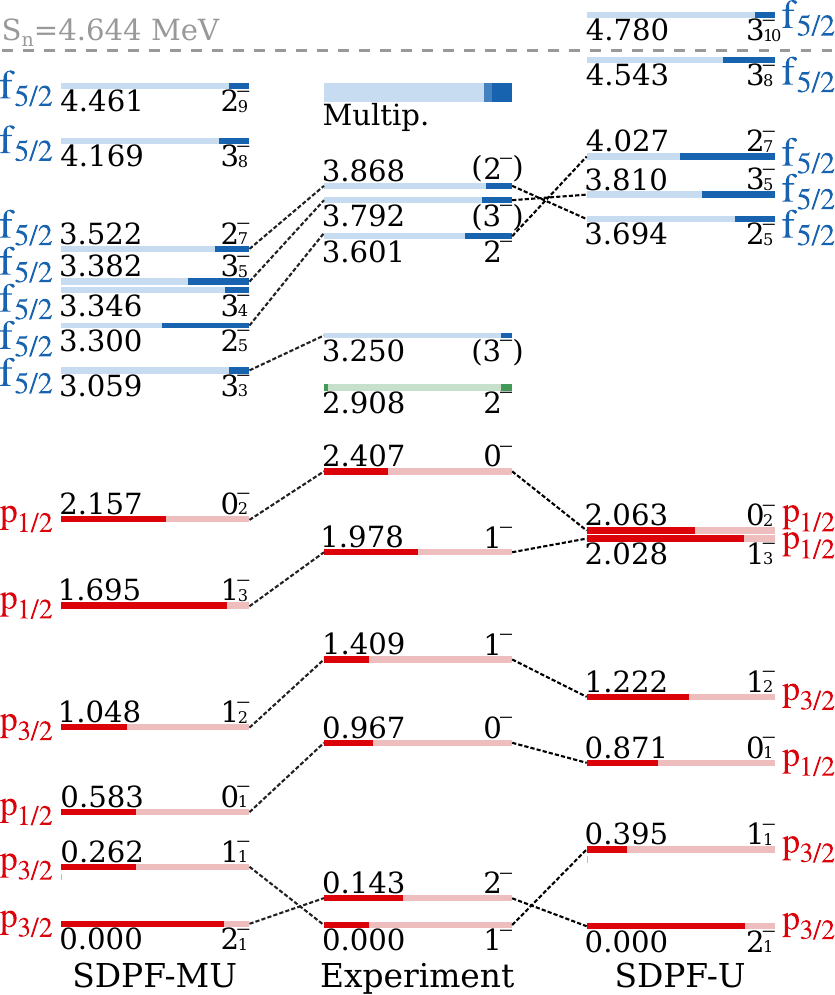}
    \caption{Partial level schemes of states directly populated by $^{47}$K($d,p$) as observed experimentally, and as predicted by shell model calculations using SDPF-MU~\cite{Otsuka2010_SDPF-MU_Vmu,Utsuno2012_SDPF-MU} and SDPF-U~\cite{Nowacki2009_SDPF-U}. For each state, the filled length of the line represents the spectroscopic factor, $S$. 
    All shell model states with $S$~$\geq$~0.1 are shown -- as such, 2$^-_2$ and 3$^-_1$ are excluded. The $p$-wave ($f$-wave) states are shown in red (blue) and fill from the left (right). The mixed state (green) fills from both sides.}
	\label{fCompareExpTheory}
\end{figure}

In order to evaluate the impact of the present work on nuclear structure around neutron number, $N$\,=\,28, we first reiterate that the state-of-the-art shell-model calculations, SDPF-MU~\cite{Otsuka2010_SDPF-MU_Vmu,Utsuno2012_SDPF-MU} and SDPF-U~\cite{Nowacki2009_SDPF-U}, fail to reproduce the 1$^{-}$ ground state of $^{48}$K, as shown in Fig.~\ref{fCompareExpTheory}. This striking feature provides a key benchmark around $N$\,=\,28 and emphasises the importance of theoretical improvements in this region, where the $\pi s_{1/2}$ and $\pi d_{3/2}$ orbitals are near-degenerate~\cite{Gade2006_OddMassKClP}. In addition, we note from Fig.~\ref{fCompareExpTheory}, and report here for the first time, that the spacing between the pure $\ell=1$ and pure $\ell=3$ states (observed to be $\sim$0.8~MeV) appears to be overestimated by SDPF-U ($\sim$1.6~MeV), and is more closely reproduced by SDPF-MU ($\sim$0.9~MeV). More quantitatively, the average energies of $\nu p_{3/2}$, $\nu p_{1/2}$ and $\nu f_{5/2}$ -- weighted according to $S$ and $(2J+1)$ -- are experimentally determined to be 0.45(2), 1.93(2) and 4.0(1)~MeV, respectively~\cite{Paxman2024_Thesis}. This compares more favourably to SDPF-MU (0.56, 1.88 and 3.87~MeV) than SDPF-U (0.66, 2.12 and 4.35~MeV). Further, the splitting between the high and low spin projection of $\nu p_{3/2}$, $\nu p_{1/2}$ and $\nu f_{5/2}$ is experimentally determined to be 0.42(2), 0.19(1) and <0.8~MeV, respectively, which again seems to favour SDPF-MU (0.76, 0.10 and 0.11~MeV) over SDPF-U (1.16, 0.44 and 0.24~MeV). Recall, however, that neither shell model calculation predicts a neutron-mixed state with both $p-$wave and $f-$wave contributions, as observed in Fig.~\ref{fCSfits}(g).

With the exception of the ground and first excited state in $^{48}$K, both shell models correctly reproduce the ordering of observed $\ell=1$ transfer levels. 
Furthermore, all excited states that are predicted to exhibit strong $p-$wave single particle strength by the shell model are observed experimentally. Interestingly, however, the experimentally deduced spectroscopic factors, $S_{\textrm{exp}}$, are considerably reduced relative to the proposed shell-model counterparts. Specifically, measured $S_{\textrm{exp}}$ values are a factor $\sim$0.5$-$0.7 lower than $S_{\textrm{MU}}$ or $S_{\textrm{U}}$, as shown in Table I. This appears to be similarly true for $\ell=3$ transfer, in spite of the fact that this analytical method should reproduce the shell model spectroscopic factors~\cite{JennyLee20percent, Matta2019_29Mg}. An explanation for the observed discrepancies between theory and experiment may arise from the fact that states in $^{48}$K with two different proton configurations coexist in the same energy regime. This is due to the near degeneracy of the \piSOneDFour{} and \piSTwoDThree{} configurations. Assuming a single step process in the $^{47}$K($d,p$) reaction, states in $^{48}$K should be populated via the \piSOneDFour{} component of their wavefunction. However, rather astonishingly, we find that the overestimation of spectroscopic factors in both shell-model calculations actually increases according to the percentage of the \piSOneDFour{} configuration included in the $^{48}$K final state wavefunction~\cite{Paxman2024_Thesis}. This is true for all states but is exemplified, most prominently, by the 1$^{-}$ ground state, for which the two shell model calculations have differing results. The SDPF-U model expects \piSOneDFour{} to be only $\sim$14\% of the 1$^{-}_1$ wavefunction, which is much smaller than the SDPF-MU prediction of $\sim$28\%; accordingly, the spectroscopic factor of SDPF-U is in full agreement with experimental data, whereas SDPF-MU over-predicts the spectroscopic factor by nearly a factor of 2. Given that both SDPF-MU and SDPF-U suffer from this possible mixing issue, it would imply that matrix elements, common to both, that mix the proton \piSOneDFour{} and \piSTwoDThree{} configurations should be adjusted.

Exploring the importance of shell-model matrix elements further, we consider the two lowest 1$^{-}$ states, which manifest from a neutron in the $p_{3/2}$ orbital coupled to the \piSOneDFour{} and \piSTwoDThree{} configurations, respectively. Both shell-model interactions predict these two configurations to mix substantially, and, indeed, this is observed in the current study. It is worth noting that, in a weak coupling model, the two configurations are degenerate, whilst, conversely, mixing pushes them apart in energy. On this point, shell-model calculations predict the first two 1$^{-}$ states to be separated by less than 1~MeV; experimentally, the separation is 1.4~MeV. This is of special significance, as it is the underestimation of 1$^{-}_1$$-$1$^{-}_2$ splitting that is responsible for the incorrect prediction of a 2$^{-}$ ground state in both SDPF-MU and SDPF-U. From an investigation of a wide range of matrix elements, we find that the splitting of 1$^-$ states is especially sensitive to the matrix element that describes the interaction of a neutron in $p_{3/2}$ orbital scattering a proton between $s_{1/2}$ and $d_{3/2}$, coupled to two units of angular momentum (usually written as $<p_{3/2}s_{1/2}|V_{int}|p_{3/2}d_{3/2}>_{J=2,T=0}$). Increasing the magnitude of this matrix element by $\sim$1~MeV reproduced the experimental 1$^{-}$ splitting for both SDPF-U and SDPF-MU. A subsequent increase in the magnitude of $<p_{3/2}s_{1/2}|V_{int}|p_{3/2}s_{1/2}>_{J=1,T=0}$ by $\sim$1~MeV allowed for a successful reproduction of the ordering and spacing of all pure $\ell=1$ states. These variations had no effect on the 0$^{-}_1$$-0^{-}_2$ splitting and a only minor effect on the 2$^{-}_1$$-2^{-}_2$  splitting -- both of which are well-reproduced by SDPF-MU and SDPF-U. Furthermore, adjustments of the two matrix elements resulted in no significant effect on the structure of the $^{47}$K ground state or $\ell=1$ excited states in $^{48}$K.  
Consequently, we conclude that the present experiment provides the ideal benchmark for calculations of cross-shell neutron-proton interactions between $\nu p_{3/2}$ and $\pi s_{1/2}$ orbitals. 

In summary, we have performed the first measurement of the $^{47}$K($d,p\gamma$)$^{48}$K transfer reaction. Nine new, bound excited states have been uniquely identified, providing a benchmark for $\pi s_{1/2}$$-\nu fp$ interactions in theoretical models. A detailed comparison with state-of-the-art shell-model calculations indicates several key discrepancies between theory and experiment. In particular, the failure of shell-model calculations to predict the 1$^{-}$ ground state of $^{48}$K is traced to incorrect mixing between the proton configurations, as evidenced by the energy gap from 1$^{-}_{1}$ to 1$^{-}_{2}$. In fact, an investigation of shell-model matrix elements reveals that an adjustment of $<p_{3/2}s_{1/2}|V_{int}|p_{3/2}d_{3/2}>_{J=2,T=0}$ and $<p_{3/2}s_{1/2}|V_{int}|p_{3/2}s_{1/2}>_{J=1,T=0}$ by $\sim$1~MeV not only correctly reproduces the 1$^-$ ground state of $^{48}$K in both SDPF-U and SDPF-MU calculations, but also the ordering and spacing of all pure $p-$wave states. Furthermore, an analysis of the spacing and splitting of $f$- and $p$-wave orbitals reveals a preference for SDPF-MU (in which the $\pi sd$$- \nu fp$ interaction is derived from first principles) over SDPF-U (in which the $\pi sd$$- \nu fp$ interaction is phenomenological). Interestingly, this is despite both interactions reproducing the single-particle states in $^{49}$Ca (at 0, 2.023 and 3.991 MeV) well, and SDPF-U being fitted, in part, to experimentally measured $^{46}$Ar(d,p) data~\cite{Gaudefroy2006_46Ardp_Original,Nowacki2009_SDPF-U}. Consequently, it is apparent that the presently determined level structure of $^{48}$K now represents a key benchmark for calculations involving interactions between $fp-$shell neutrons and valence protons in the $s_{1/2}$ orbital. Such benchmarks of exotic inter-orbital interactions are critical for extrapolations to more exotic systems and the current work provides a crucial step toward accurate predictions of nuclei, such as $^{44}$S~\cite{Glasmacher1997_44S-Collective}, around the $N=28$ `island of inversion' -- a region of the nuclear chart in which configurations with
particle-hole excitations intrude to be lower in energy than orbitals filled in the conventional order. Finally, experimentally determined spectroscopic factors in $^{48}$K are found to be systematically smaller than those predicted by theory, which may be due to an overestimation by shell-model calculations of the fraction of \piSOneDFour{} in the final state wavefunctions. This echoes the difficulty with shell-model calculations for neighbouring $^{47}$Ar, populated by $^{46}$Ar($d,p$)~\cite{Gaudefroy2006_46Ardp_Original}, indicating that the key limitation in our description of light nuclei around $N=28$ is the poor understanding of the proton component, which the present $^{47}$K(d,p) reaction has exposed to scrutiny. 
Thus, we strongly encourage that further studies of single-nucleon transfer on exotic $N = 28$ isotones, such as $^{45}$Cl and $^{43}$P, be performed, in order to further explore the influence of near-degenerate proton configurations on nuclear structure in this region.

\begin{acknowledgments}
We acknowledge the GANIL facility for provision of heavy-ion beams, and we thank J. Goupil, G. Fremont, L. M{\'e}nager, and A. Giret for assistance in using the G1 beam line and its instrumentation. We acknowledge
the AGATA collaboration for the use of the spectrometer. This work was supported by the Science Technology Facilities Council (United Kingdom) under the grants ST/N002636/1, ST/P005314/1 and ST/V001108/1. This work was also partially funded by MICIU MCIN/AEI/10.13039/501100011033, Spain with grants PID2020-118265GB-C42, PRTR-C17.I01, Generalitat Valenciana, Spain with grant CIPROM/2022/54, ASFAE/2022/031, CIAPOS/2021/114 and by the EU NextGenerationEU funds. 
Additional funcing from Spanish grant PID2021-127711NB-100. This work was partially supported by the U.S. Department of Energy, Office of Science, Office of Nuclear Physics, under contract number DE-AC02-06CH11357. Experimental data and analytical codes can be found at Refs.~\cite{RawData, AnalyticalCode, ProcessedData}
\end{acknowledgments}

\bibliographystyle{apsrev4-1}
\bibliography{biblio}

\end{document}